\documentclass{aa}  

\usepackage{graphicx}
\usepackage{amsmath}
\usepackage{amssymb}
\usepackage{threeparttable}
\usepackage[final]{microtype}
\usepackage{url}
\usepackage{txfonts}
\usepackage{hyperref}
\def \her {\object{\mbox{Her X-1}}\xspace}
\def \sb  {\emph{Swift}/BAT\xspace}

\begin{document} 

\title{Swift/BAT measurements of the cyclotron line energy decay in
  the accreting neutron star  
  \mbox{Her X-1}: indication of an evolution of the magnetic field?
  } 

\author{D.~Klochkov\inst{1} \and
        R.~Staubert\inst{1} \and
        K.~Postnov\inst{2}  \and
        J.~Wilms\inst{3}    \and
        R.\,E.~Rothschild\inst{4}    \and
        A.~Santangelo\inst{1}
      }

\institute{Institut f\"ur Astronomie und Astrophysik, Universit\"at
           T\"ubingen (IAAT), Sand 1, 72076 T\"ubingen, Germany 
           \and
           Moscow M.V. Lomonosov State University, Faculty of Physics
           and Sternberg Astronomical Institute, 199992, Moscow, Russia 
           \and
           Dr. Remeis-Sternwarte \& ECAP, Astronomisches Institut der
           Universit\"at Erlangen-N\"urnberg, Sternwartstr. 7, 96049
           Bamberg, Germany 
           \and
           Center for Astrophysics and Space Sciences, University of
           California, San Diego, 9500 Gilman Dr., La Jolla, CA
           92093-0424, USA 
         }

\date{Received ****; accepted ****}

\abstract
{
  The magnetic field is a crucial ingredient of neutron stars. It governs
  the physics of accretion and of the resulting high-energy
  emission in accreting pulsars. Studies of the cyclotron 
  resonant scattering features (CRSFs) seen as absorption lines in the
  X-ray spectra of the pulsars permit direct
  measuremets of the field strength.
}
{ 
  From an analysis of a number of pointed observations with different
instruments, the energy of CRSF, $E_{\rm cyc}$, has recently
been found to decay
  in \her, which is one of the best-studied accreting pulsars. 
  We present our analysis of a 
  homogeneous and almost uninterrupted monitoring of the line energy 
  with \sb.
}
{ 
  We analyzed the archival \sb observations of \her from 2005
  to 2014. The data were used to measure the CRSF energy
  averaged over several months.
}
{
  The analysis confirms the long-term decay of the
  line energy. The downward trend is highly significant and consistent
  with the trend measured with the pointed observations: 
  ${\rm d}E_{\rm cyc}/{\rm d}t\sim -0.3$\,keV per year.
}
{
  The decay of $E_{\rm cyc}$ either indicates a local evolution of the
  magnetic field 
  structure in the polar regions of the neutron star or a geometrical
  displacement of the line-forming region due to long-term changes in
  the structure of the X-ray emitting region. The shortness of the
  observed timescale of the decay, $-E_{\rm cyc}/\dot E_{\rm cyc}\sim
  100$\,yr, suggests that trend reversals and/or jumps of the line
  energy might be observed in the future.
}

\keywords{neutron stars -- accretion -- magnetic field}
\titlerunning{Evolution of magnetic field in Her X-1}
\maketitle

%
\section{Introduction}
\label{sec:intro}

Accreting magnetized neutron stars are among the brightest Galactic
X-ray sources, with luminosities reaching
$10^{38}$--$10^{39}$~erg~s$^{-1}$. They are powered by the
gravitational energy of matter that is supplied by the binary companion star
and accreted by the compact neutron star (NS). The magnetic
field of the NS is believed to have a roughly dipole
configuration, with a field strength at the star surface of 
$B\sim 10^{12}-10^{13}$\,G. In the vicinity of the NS, the
accretion flow consisting of heated ionized gas is channeled by the
magnetic field lines toward the polar caps of the NS. In these
regions, matter arrives at the stellar surface at a velocity of $\sim$0.5$c,$
causing energy release in form of X-rays. The rotation of the
NS causes a periodic modulation of the observed flux -- X-ray
pulsations. The sources are therefore also referred to as X-ray
binary pulsars (XBP) or accreting pulsars. 

X-ray binary pulsars comprise the second most numerous observed population
of NSs after radio pulsars: $\sim$200 XBPs are currently
known\footnote{See, e.g., the on-line catalogue by M.~Orlandini: 
\url{http://www.iasfbo.inaf.it/~mauro/pulsar_list.html}}.
An important advantage of these objects is that
their magnetic field strength can be measured through
observing \emph{\textup{cyclotron resonance scattering features}} (CRSFs).
These features appear as absorption lines (cyclotron lines), caused by the 
resonant scattering of photons off the electrons in Landau levels
\citep[e.g.,][]{Truemper:etal:78,Isenberg:etal:98,Schoenherr:etal:07}. The
energy  $E_{\rm cyc}$ of the fundamental line and the spacing between
the harmonics are directly proportional to the field strength: 
$E_{\rm cyc}\simeq 12B_{12}(1+z)^{-1}$~keV, where $B_{12}$ is the magnetic
field strength in units of $10^{12}$\,G and $z$ is
the gravitational redshift at the region where the line is formed.
The energies of the cyclotron lines in XBPs range from $\sim$10 to
$\sim$100\,keV \citep[e.g.,][]{Caballero:Wilms:12}.

\her is the first XBP where a CRSF was discovered and interpreted as 
such almost forty years ago \citep{Truemper:etal:78}. 
It is interesting to note that the feature was originally
interpreted as an emission line at $\sim$58\,keV: part of the
X-ray continuum on the high-energy side of the absorption feature was
modeled with an emission component. The subsequent observations and
analyses have shown, however, that the feature is an absorption line with a
centroid energy of $\sim$35\,keV \citep[e.g.,][and references
therein]{Gruber:etal:01}. Among more than a dozen confirmed CRSF
sources, \her possesses one of the most prominent 
cyclotron lines whose characteristics can be reliably measured with
shorter observations than in other XBPs at a similar flux level. 
The source shows a regular pattern of \emph{\textup{on}} and \emph{\textup{\textup{of}f}} states
repeating with a period of $\sim$35\,d. 
One 35\,d cycle includes two \emph{\textup{on}} states: the \emph{\textup{main-on}}
characterized by a peak flux of up to 300\,mCrab, and the
\emph{\textup{short-on}} when the flux is$\text{ about }$five times lower.
The periodicity is believed to be caused by the warped 
accretion disk precessing with a $\sim$35\,d period \citep[e.g.,][and
references therein]{Klochkov:etal:06}. 
In most cases, only the data from main-ons have sufficient quality for
the analysis of the cyclotron line. Thanks to its almost persistent
nature and near regularity of the on-states, \her has repeatedly been
observed with almost every hard 
X-ray instrument operated since the discovery of the source. As a result, 
an unprecedentedly long record of CRSF measurements in a single
pulsar has been accumulated. For a detailed description of the
observational data base of the cyclotron line measurements in \her, we
refer to the recent work by \citealt{Staubert:etal:14}
(hereafter, Paper~I).  

The CRSF centroid energy $E_{\rm cyc}$ in \her does not remain constant. 
In addition to the regular variability with the
1.24\,s pulse phase of the NS, which is very common in XBPs,
the cyclotron line energy in \her is found to depend on the
source luminosity \citep{Staubert:etal:07,Klochkov:etal:11}. The
luminosity $L_{\rm X}$ of \her, which reflects the mass accretion rate $\dot M,$
changes stochastically on a long timescale (from one 35\,d cycle to
the next) as well as on the timescale of individual 1.24\,s pulse
cycles (i.e., from one pulse cycle to the next, called \emph{\textup{pulse-to-pulse}}
variability). By comparing the CRSF energies at different
luminosities, it was found that $E_{\rm cyc}$ is directly
proportional to the X-ray luminosity on both timescales.

\begin{figure}
\centering
\resizebox{1.1\hsize}{!}{\includegraphics{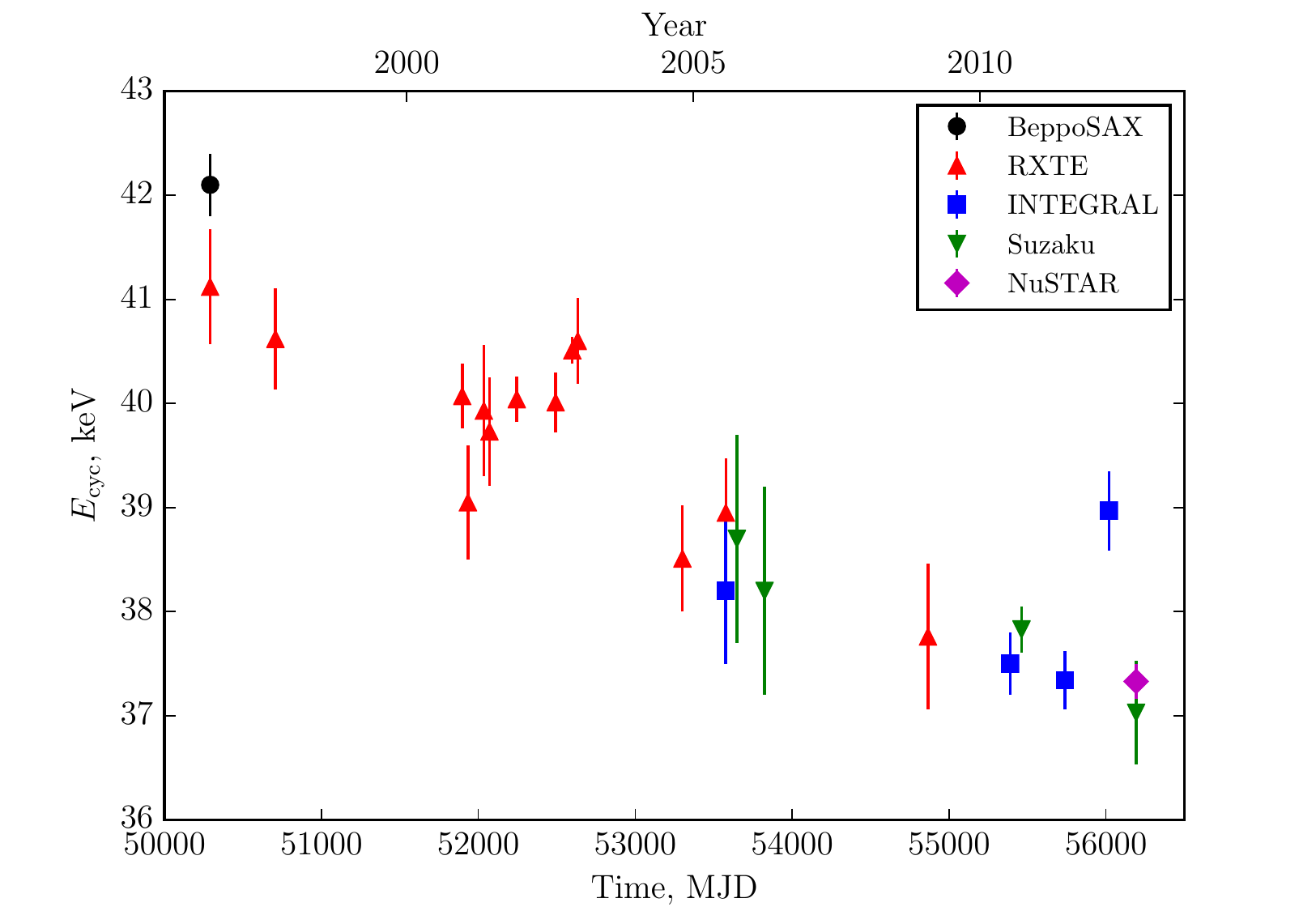}}
\caption{Centroid energy $E_{\rm cyc}$ of the CRSF in \her
  measured with different instruments since 1995 \citep[see][for
  details]{Staubert:etal:14}. 
}
\label{fig:pointed}
\end{figure}

In addition to and independently of the $E_{\rm cyc}-L_{\rm X}$
relation, it was demonstrated in Paper~I that the line centroid
energy has been significantly evolving with time since the beginning of the
observations. Specifically, it is shown that $E_{\rm cyc}$ has been
systematically decreasing since $\sim$1995 from $\sim$41 to
$\sim$37\,keV, that is, by $\sim$10\% in total. In Fig.\,\ref{fig:pointed}, we
plot the line energy measured in pointed observations (all performed
in main-on state of the source) as a function of time since 1995. 
See Paper~I for a discussion of the $E_{\rm cyc}$ measurements
obtained before 1995 for which the original data are no longer available for 
re-analysis. 

A significant downward trend of the CRSF energy with time can be clearly
seen in Fig.\,\ref{fig:pointed}. The X-ray luminosity $L_{\rm X}$
of \her (characterized by the maximum flux during a main-on) changes
stochastically between the observations without 
showing any long-term trend. Since $E_{\rm cyc}$ is proportional to
$L_{\rm X}$, the measured energies of the cyclotron line are affected
by the varying luminosity of the source. To take this dependence into
account, in Paper~I we modeled the measured CRSF energy by a linear
function of both time and luminosity $E_{\rm cyc} = F(t,L_{\rm
  X})$. The time dependence of the line energy was found to be
characterized by the time derivative 
${\rm d}E_{\rm cyc}/{\rm d}t = -0.26\pm0.01$\,keV/yr, indicating a
significant trend.

In this work, we make use of the publicly available archival data taken
with the Burst Alert Telescope (BAT, \citealt{Barthelmy:etal:06}) -- a
hard X-ray detector onboard the 
\emph{Swift}  orbital observatory \citep{Gehrels:etal:04}. The BAT instrument
has a huge field of view of 1.4 steradian and is designed to
provide triggers and accurate positions for gamma-ray bursts. 
While searching for bursts and other transient sources, \textsl{BAT} 
points at different locations in the sky, thus performing 
an all-sky monitoring in hard X-rays \citep[see, e.g.,][]{Krimm:etal:13}.
Here, we analyze the BAT data taken on \her since the launch
of the mission, from 2005 to the end of 2014. The BAT data have some
important advantages compared to the available data from the pointed
observations. First, the regular visits of the source by the
instrument provide a nearly homogeneously spaced set of data without
long time gaps. Second, the measurements taken with the same instrument
can be safely compared with each other. Third, in each single measurement
of $E_{\rm cyc}$, we summed the data from several main-ons
($\gtrsim$five). Thus, the variation of the cyclotron
line energy due to stochastic variability of $L_{\rm X}$ from one
main-on state to the next is expected to be mitigated ($E_{\rm cyc}$ is
proportional to $L_{\rm X}$, see above). 

\section{Reduction and analysis of the BAT survey data}
\label{sec:processing}

\begin{figure}
\centering
\resizebox{\hsize}{!}{\includegraphics{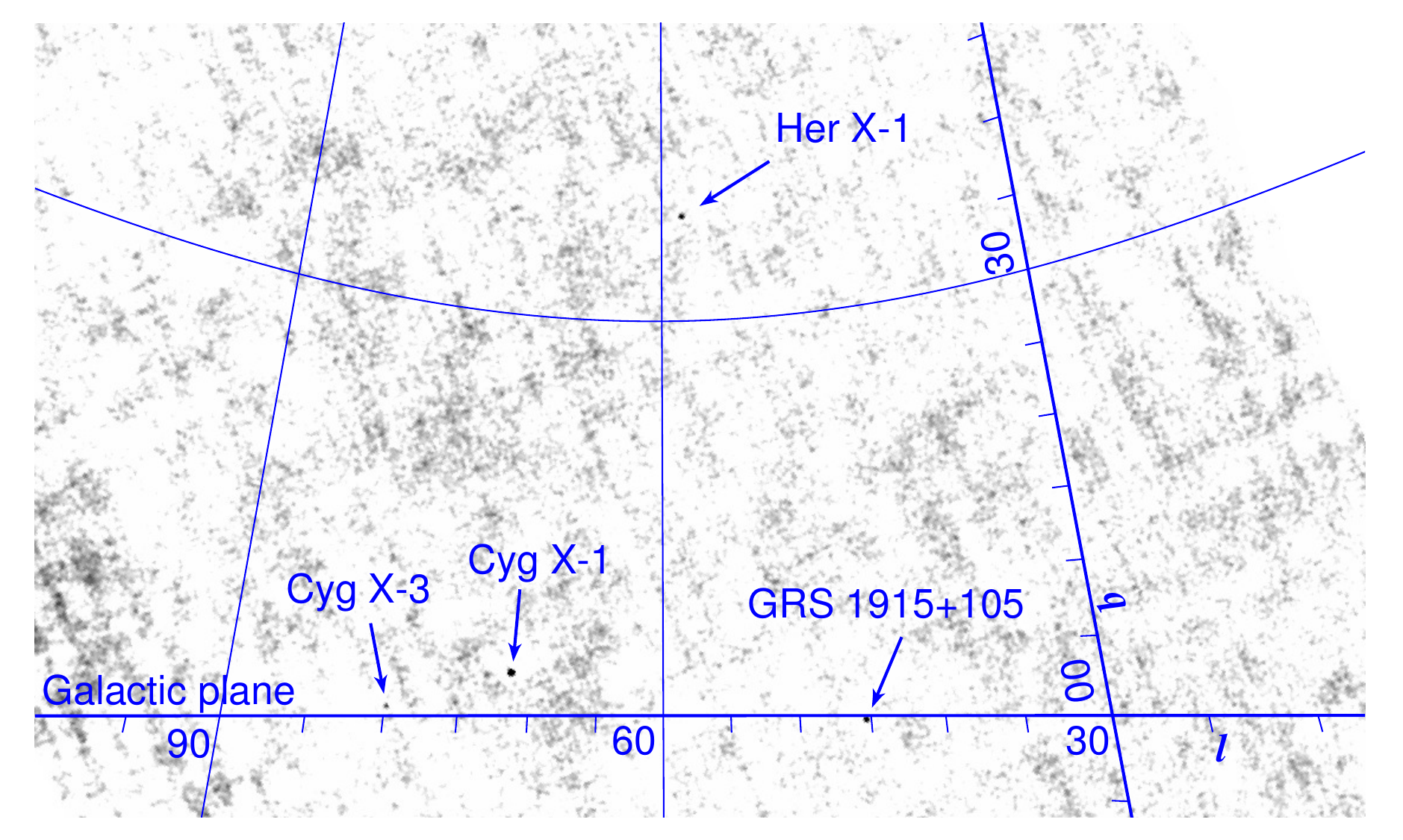}}
\caption{Example of a sky image in Galactic coordinates taken with
  \sb in one observation 
  (in this case, $\text{about }$one\,ks). \her and a few other
  detected hard X-ray sources are indicated.
}
\label{fig:skyima}
\end{figure}

The \sb data used in this work are available through 
HEASARC\footnote{\url{http://heasarc.gsfc.nasa.gov/cgi-bin/W3Browse/swift.pl}}.
The BAT instrument is sensitive in the photon energy range
$\sim$15--150\,keV and is, therefore, well suited for the study of the
$\sim$40\,keV feature in \her. Most of the BAT data are stored in the
form of detector plane maps (histograms) accumulated over a five-minute
exposure time. We collected such maps taken during the main-on states of
\her (35\,d phases $\sim$0.0--0.25, where phase zero corresponds to
the time of X-ray turn-on -- the sharp transition from the off- to the
main-on state) when the source is in the BAT
field of view starting from 2005  
(start of \emph{Swift} operation) to the end of 2014 (the time of
the work on the publication). 
We then used the maps to construct the sky images with the help of the
tools \emph{\textup{\textit{batbinevt}}} and \emph{batfftimage} from the
HEASoft~6.16\footnote{\url{http://heasarc.nasa.gov/lheasoft}} package.
For further analysis, we only took the data where \her is clearly
detected in the image. One such image accumulated in an observation of
about 1\,ks is shown in Fig.\,\ref{fig:skyima}.  

\begin{figure}
\centering
\resizebox{\hsize}{!}{\includegraphics{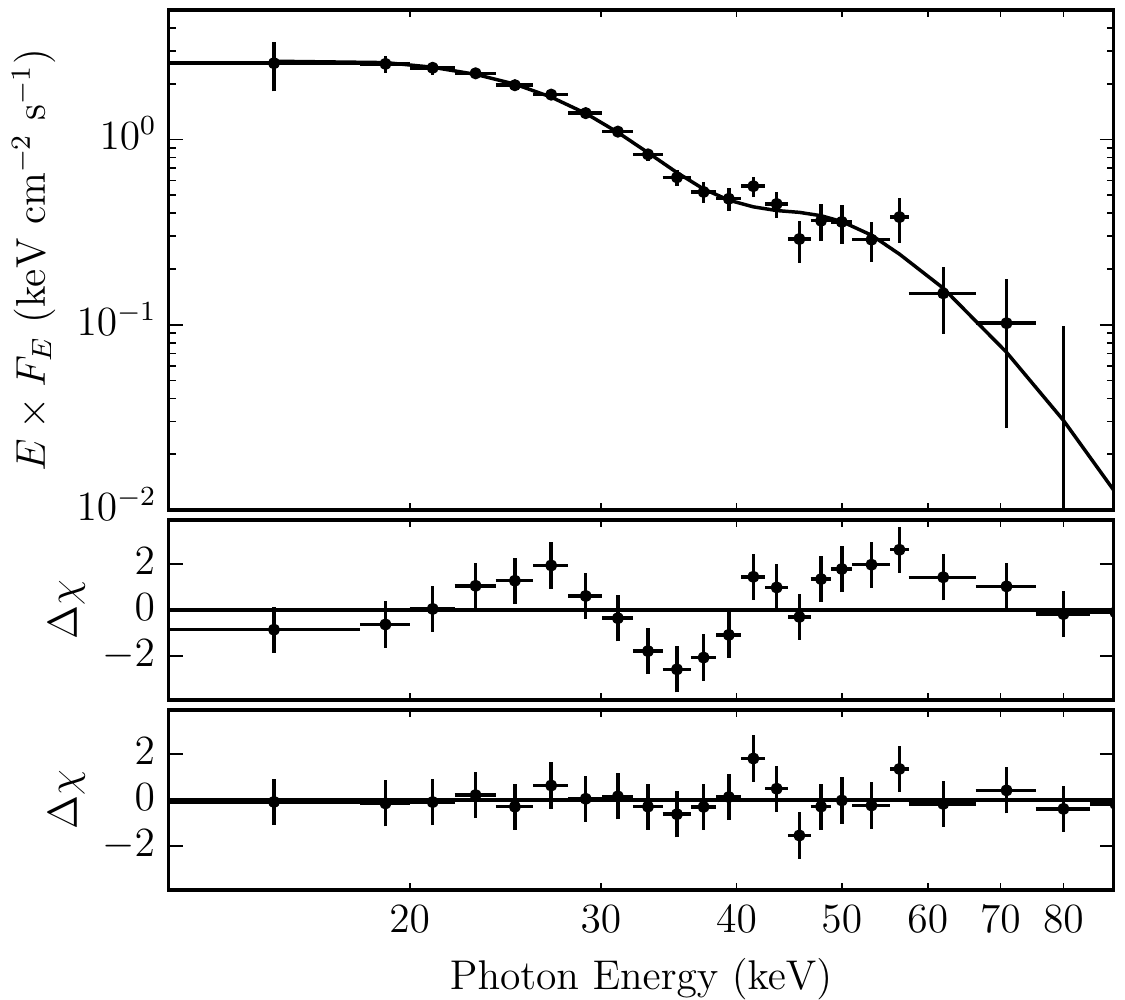}}
\caption{Example of a \sb spectrum of \her accumulated during 
an observation of $\text{about}$15\,ks fitted with a power-law-cutoff model with a
Gaussian absorption line representing the CRSF (top) and the residuals
after fitting the spectrum without the Gaussian line (middle) and
including the line (bottom). 
}
\label{fig:spe}
\end{figure}

Detector plane histograms of BAT also contain the spectral information
of the detected X-rays with the photon energy binning determined by
80 pre-defined energy channels. We used this information to extract
spectra of \her using the multipurpose task \emph{batbinevt}
mentioned above. To correct the BAT energy scale for the nonlinear
behavior of the detector electronics and for the detector-dependent
offsets not accounted for by the on-board calibration, we used the
tool \emph{baterebin} and the available data base of the
gain/offset maps of the 
detector. The tool applies a detector-dependent energy shift to the
data. According to the instrument team, the accuracy of the corrected
energy scale is $\pm$0.1\,keV based on the spectrum of the on-board
$^{241}$Am calibration 
source\footnote{\url{http://swift.gsfc.nasa.gov/analysis/bat_digest.html}}.
To the resulting \her spectra we added an energy dependent systematic
uncertainty of $\geq$4\% using the tool \emph{batphasyserr,} as
suggested by the instrument team. An example of the extracted BAT
spectrum of \her accumulated in an exposure time of $\sim$15\,ks is presented in 
Fig.\,\ref{fig:spe}. 
To model the obtained BAT spectra, we used
the XSPEC \emph{cutoffpl$\times$gabs} model (power law with an
exponential cutoff modified by an absorption line of a
Gaussian optical depth profile). The power-law-cutoff model
\emph{cutoffpl} is the simplest continuum model of those that
are commonly used to fit hard X-ray continua of XBPs. The Gaussian absorption line
model \emph{gabs} represents the cyclotron absorption line similarly
to Paper~I and to the most of the recent works on X-ray spectroscopy of \her
\citep[e.g.,][]{Staubert:etal:07,Klochkov:etal:08b,Fuerst:etal:14c}.
The solid line in the top panel of Fig.\,\ref{fig:spe} shows the
best fit of the spectrum with the described spectral model.
The model provides stable fits with all our BAT spectra and results in line
energies that are consistent with those measured in the pointed observations
using other instruments, as shown in the next section. 

To achieve a data quality sufficient for measuring the CRSF centroid
energy with uncertainties similar to those of the pointed
observations (Fig.\,\ref{fig:pointed}), we summed BAT spectra of
the source from four to ten adjacent main-on states. 
Although the BAT data coverage of a particular source is
roughly homogeneous on a timescale of several hundred days, a
significant clustering occurs on shorter timescales. We
therefore used such a variable (adaptive) size of the time
bins to maintain approximately the same accuracy for each measurement
and to use as many BAT datasets useful for the spectral extraction
as possible. Each of our BAT
measurements of $E_{\rm cyc}$ thus corresponds to a time span of
a few months. The CRSF centroid energies measured in this way using the
data from the beginning of the \emph{Swift} operation are
provided in Table\,\ref{tab:ecyc}. 

\begin{table}
  \centering
  \renewcommand{\arraystretch}{1.3}
  \caption{ CRSF centroid energy $E_{\rm cyc}$ measured with \sb in
  \her since the beginning of the \emph{Swift} operation. The quoted
  uncertainties are at 1$\sigma$ confidence level.} 
  \label{tab:ecyc}
  \begin{tabular}{l l l}
    \hline\hline
    MJD start & MJD end & $E_{\rm cyc}$ [keV]\\
    \hline
              &       &     \\
    53440     & 53580 & $39.37_{-0.38}^{+0.36}$\\
    53610     & 53750 & $39.53_{-0.88}^{+0.89}$\\       
    53790     & 54030 & $39.62_{-0.19}^{+0.19}$\\
    54070     & 54210 & $39.26_{-0.46}^{+0.43}$\\
    54240     & 54450 & $38.80_{-0.30}^{+0.30}$\\       
    54480     & 54590 & $37.73_{-0.35}^{+0.49}$\\
    54620     & 54970 & $38.57_{-0.63}^{+0.68}$\\
    55010     & 55150 & $38.71_{-0.54}^{+0.26}$\\
    55180     & 55320 & $38.33_{-0.42}^{+0.25}$\\       
    55360     & 55530 & $36.84_{-0.40}^{+0.43}$\\
    55570     & 55740 & $37.00_{-0.54}^{+0.59}$\\
    55780     & 55920 & $36.89_{-0.24}^{+0.25}$\\
    55950     & 56090 & $36.32_{-0.44}^{+0.46}$\\
    56120     & 56260 & $36.35_{-0.40}^{+0.42}$\\
    56300     & 56470 & $37.80_{-0.60}^{+0.63}$\\
    56510     & 56610 & $37.45_{-0.33}^{+0.34}$\\
    56650     & 56790 & $36.87_{-0.55}^{+0.58}$\\
    56820     & 57030 & $37.09_{-0.23}^{+0.23}$\\
    \hline
  \end{tabular} 
\end{table}

\section{Secular decay of the CRSF energy}
\label{sec:decay}

\begin{figure}
\centering
\resizebox{1.1\hsize}{!}{\includegraphics{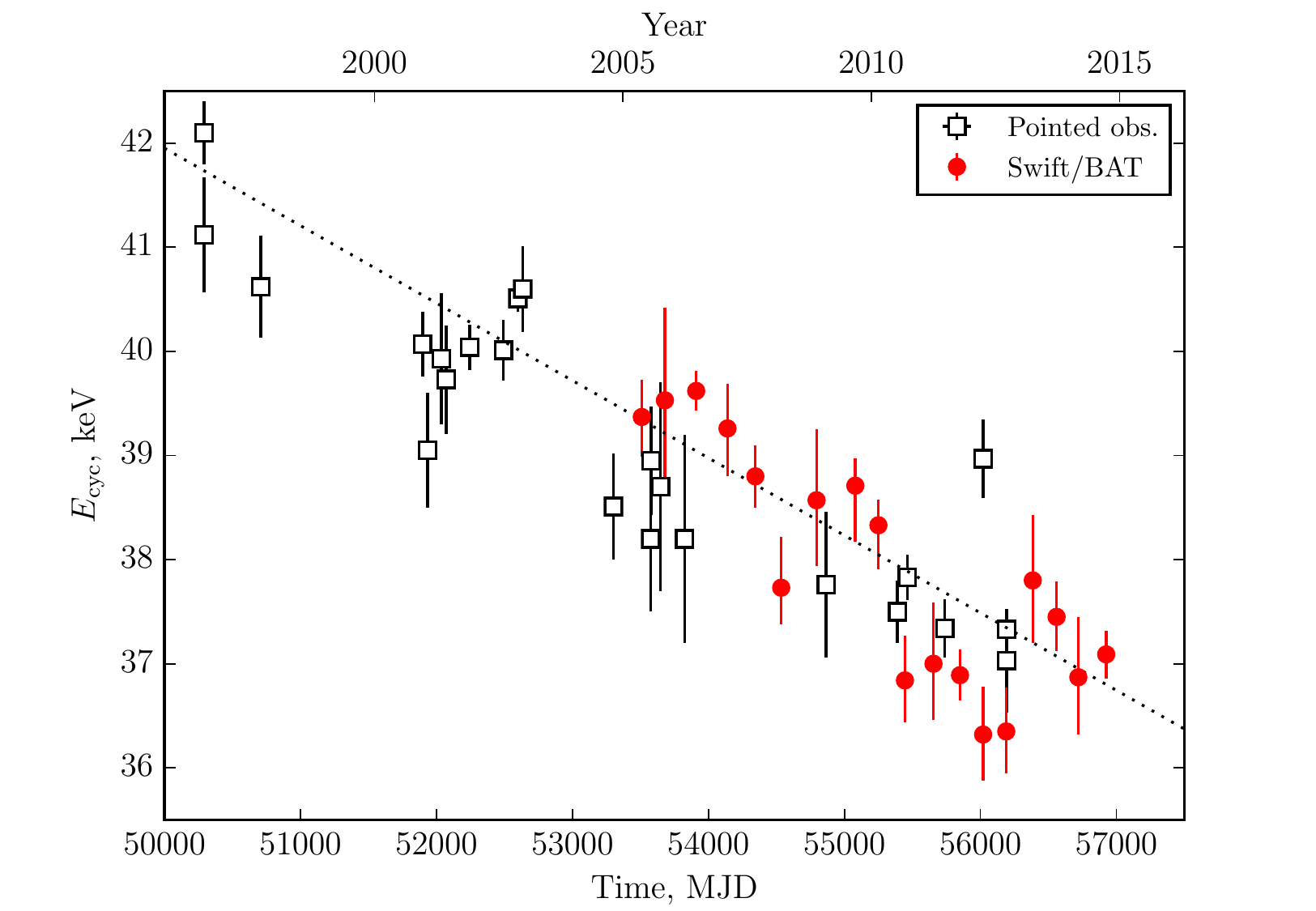}}
\caption{ CRSF centroid energy $E_{\rm cyc}$ in \her measured with
  pointed observations (black squares, same as in
  Fig.\,\ref{fig:pointed}) and our \sb measurements (red circles). The
  dotted line indicates a linear fit to the data from only the pointed
  observations. The error bars in all cases indicate the
  uncertainties at 1$\sigma$ confidence level.
} 
\label{fig:ecyc}
\end{figure}

The presented cyclotron line energies measured with \sb demonstrate a
clear downward trend very similar to that indicated by the pointed
observations. In Fig.\,\ref{fig:ecyc}, the \sb measurements presented
in Table\,\ref{tab:ecyc} are plotted as red
filled circles together with the measurements from the pointed
observations (rest of the data points, identical to
Fig.\,\ref{fig:pointed}). To demonstrate that the BAT values generally
follow the trend indicated by the pointed 
observations, we added a linear fit to the pointed observations alone
shown by the dotted line. Some deviations between the BAT and the
pointed measurements reaching a few standard deviations are present
and might be due to unknown BAT systematics and/or due to imperfect
cross-calibration of the BAT spectral response with respect to the
other instruments. The variability of $E_{\rm cyc}$ on a timescale
of months due to the $E_{\rm cyc}-L_{\rm X}$ dependence might also
contribute to the deviations because the BAT measurements cover much
longer time spans than the pointed observations.

Despite the deviations, the long-term evolution of $E_{\rm cyc}$
indicated by the BAT data is consistent with that exhibited by the
pointed observations. To demonstrate this quantitatively, we calculated
the slope of the $E_{\rm cyc}(t)$ dependence from a linear fit to the data.
A fit to the BAT data alone yields 
${\rm d}E_{\rm cyc}/{\rm d}t=-0.32(3)$\,keV/yr, which is consistent
(within $\sim$2$\sigma$) with the values $-0.26(1)$ and $-0.28(1)$
keV/yr reported in Paper~I for the pointed observation with and
without taking the $E_{\rm cyc}-L_{\rm X}$ dependence into account,
respectively. The significance of the linear trend in the BAT data
alone from a linear correlation analysis is characterized by a two-sided
null hypothesis probability (to obtain the correlation/trend by chance) of
$\sim$10$^{-5}$.

\section{Summary and discussion}
\label{sec:discussion}
The analyzed \sb observations of \her clearly confirm the long-term
decrease of the cyclotron line centroid energy $E_{\rm cyc}$ with a
rate of $\sim$0.3\,keV per year reported in Paper~I based on the pointed
observations with different instruments. 
The BAT data set provides a more continuous and homogeneous although
less accurate and somewhat shorter (starting only from 2005)
monitoring of the line energy than do the available 
pointed observations. As mentioned in the introduction, each BAT data
point corresponds to measurements over several adjacent on-states of
the source such that $E_{\rm cyc}$ variations from one on-state to
the next (related to the variation of the luminosity) is partially
averaged out. The data also provide an extension of the
monitoring after the latest available pointed observations. We plan to
continue our analysis of the upcoming BAT data and follow the evolution of the
line energy in the future, also with more pointed observations.

In Paper~I, we discussed a number of physical effects that might cause
the observed decay of the CRSF energy. They are mostly related either
to a possible local evolution of the $B$-field structure in/around the
polar regions of the NS driven by the accumulation and
re-distribution of the accreted plasma or to the long-term changes in the
physical dimensions of the emitting region, leading to a geometric
displacement of the line-forming region in the inhomogeneous
magnetic field. 
As \citet{Staubert:14} has pointed out, the observed effect may
be due to a slight imbalance between the rate on which matter is
accreted (gained) and the rate at which matter is lost from
the accretion mound -  either by leaking to larger areas of the
NS surface or by incorporation into the NS
crust. 

The evolution of the global dipole magnetic field of
the neutron star, generally assumed to take place in accreting pulsars,
must be characterized by the timescales of $\gtrsim$10$^{6}$\,yr
\citep[e.g.,][]{Bhattacharya:etal:92}, that is, much longer 
than that of the observed variation. Such a global evolution thus cannot
explain the observed decay of $E_{\rm cyc}$, whose characteristic
time scale is $-E_{\rm cyc}/\dot E_{\rm cyc}\sim 100$\,yr.

The local changes of the magnetic field, that are for instance due to
accumulation and spreading of accreted
matter in the magnetically confined accretion mound,
can indeed occur on much shorter timescales.
In Paper~I, we argued that the timescale for the Ohmic decay
$\tau_{\rm Ohmic} = 4\pi R^2\sigma/c^2$ for a small region like an NS polar cap
with $R\lesssim 1$\,km can be as short as $\sim$10$^2$\,yr. 
If the field evolution is driven by the crustal Hall effect, the
charactreristic timescale should be 
$\tau_{\rm Hall}\sim 5\times 10^8(R_5^2/B_{12})(\rho/\rho_{\rm
  nuc})$\,yr \citep[e.g.,][]{Goldreich:Reisenegger:92}, which for a
characteristic size of the polar cap $R\sim 1$\,km and a crustal density 
$\rho\sim 10^{11}$\,g~cm$^{-3}$ gives $\sim$10$^{5}$\,yr -- much too
long for the observed changes. If we assume, however, that the
footprint of the accretion stream on the NS surface has a shape of thin arcs 
\citep[e.g.,][]{Postnov:etal:13} with characteristic widths $R<100$\,m,
$\tau_{\rm Hall}$ falls below 1000\,yr and approaches the observed
timescale of the cyclotron line energy decay.
On the other hand, the detailed calculations
by \citet{Payne:Melatos:04}, 
although neglecting the Hall effect and elastic stresses, 
have shown that the NS needs to accumulate $\sim$10$^{-5}\,M_\odot$
for the magnetic field to be significantly reducued. For Her X-1
accreting at a rate $\dot M\sim 10^{-9}\,M_\odot$\,yr$^{-1}$, this
leads to a corresponding timescale of $\sim$10$^4$\,yr, two orders of
magnitude longer than the observed one.
We note that a relatively short timescale of 100--1000\,yr
for the $B$-field evolution depending on the equation of state of the
NS crust has recently been obtained by
\citet{Priymak:etal:14}, who ignored the Hall drift and Ohmic
diffusion, however. It is clear that more sophisticated MHD calculations are
necessary to investigate the details of the $B$-field evolution in
accreting pulsars on the observed timescale. At the current stage,
we can only argue that the observed temporal decay of the CRSF energy
in Her X-1 is most likely associated with a local effect in the
vicinity of the NS polar cap. If this is true, it is expected that the decay
should stop after some time, probably with an abrupt jump of the local  
magnetic field back to the unperturbed value. 
As discussed in Paper~I, such a jump might have
indeed been observed in the early 1990s.
However, the relatively sparse
time coverage of \her by X-ray observations did not allow
following the event in detail.

The spreading of accreted matter and the associated evolution
of the local $B$-field might lead to a complicated configuration of
the magnetic field in the polar regions of the 
NS with higher multipole components. Indeed, such a configuration
seems to be indicated by complex emitting regions necessary to model
the pulse profile evolution over the 35\,d period of the source
\citep{Postnov:etal:13}.

We conclude that the long-term monitoring of accreting
X-ray pulsars may give additional clues on the physical processes in
the neutron star crusts through observations of time-dependent
behavior of the CRSF energy. 


\begin{acknowledgements}
The research is supported by the joint DFG grant KL~2734/2-1 and
Wi~1860~11-1 and the RFBR grant 14-02-91345. 

This research has made use of data and/or software provided by the
High Energy Astrophysics Science Archive Research Center (HEASARC),
which is a service of the Astrophysics Science Division at NASA/GSFC
and the High Energy Astrophysics Division of the Smithsonian
Astrophysical Observatory. 

We thank the referee, Ulrich R.M.E. Geppert, for his
comments and suggestions that improved the the manuscript. 

DK thanks Valery Suleimanov for useful discussions.

\end{acknowledgements}


\bibliographystyle{aa}
\bibliography{refs}

\end{document}